\documentclass[natbib]{svjour3tmp}                     
\smartqed  
\usepackage{graphicx}
\usepackage{aps-bibstyle}  
\usepackage{color}
\usepackage{natbib}
\usepackage{mathpazo,bm}
\usepackage{hyperref}
\PassOptionsToPackage{pdfmark}{hyperref}\RequirePackage{hyperref}
\newcommand{\txc}{\textcolor}

\newcommand{\kms}{\,{\rm km\,s^{-1}}}
\begin{document}

  \title{Galaxies in box: A simulated view of the interstellar medium}
  
    \subtitle{}


  \author{Frederick A. Gent }


    \institute{F. A. Gent \at
                Newcaslte University \\
                Tel.: +44-191-2228586\\
                Fax: +44-191-2228020\\
                \email{f.a.gent@ncl.ac.uk}             \\
               \emph{School of Mathematics and Statistics,
                 Herschel Building,
                 Newcastle University,
                 Newcastle upon Tyne, 
                 NE1 7RU, UK}
              }

    \date{Received: 13 December 2010 / Accepted: 19 April 2011}

  \maketitle

  \begin{abstract}
     We review progress in the development of physically realistic 
     three dimensional simulated models of the galaxy. We consider the 
     scales from star forming molecular clouds to the full spiral disc.
     Models are computed using hydrodynamic (HD) or magnetohydrodynamic
     (MHD) equations and may include cosmic ray or tracer particles.
     The range of dynamical scales between the full galaxy structure
     and the turbulent scales of supernova (SN) explosions and
     even cloud collapse to form stars, make it impossible with current 
     computing tools and resources to resolve all of these in one model.  
     We therefore consider a hierarchy of models and how they can be
     related to enhance our understanding of the complete galaxy.
  \end{abstract}

  \section{Introduction}
    \label{section:intro}
    Modelling numerically any astrophysical phenomena in general has proven
challenging, even taking into consideration some considerable
simplifications. However with the improvements in computing from the late
nineties onwards reasonable 3-dimensional approximations of galactic features   have been developed.

    The aim of this section of the chapter is to briefly review progress
and to summarise some key findings to date. To impose some finite limit
on the scope of the review we restrict our attention to three dimensional
models, which include realistic parameters on galactic scales. Please
attribute any omissions as a reflection of the ignorance of this author and
not on the value of the work.

    We present a roughly chronological review of models and their methods,
objectives and results. This approach evolves out of the earlier 2D study
by \citet*{Rosen93} which investigated the hydrodynamical turbulent
structure of the ISM using stellar heating and star formation.

    \citet*{Tomisaka98} attempted to model multiple supernovae (SNe) in 3D
using a super-bubble approximation of in a stratified ISM with magnetic
field. We summarise in Table~\ref{table:sim} some key features of the models,
which followed subsequently. 

  \section{Review of galaxy simulations in 3D}
    \label{section:1}

  In the following sections we summarise the work of several groups, who have 
published in some cases several papers over the last decade. For brevity we
have cited the more recent publications, whilst including material from
earlier papers. Please refer to the cited papers for the full list of
citations.

\begin{table}
    \rotatebox{90}{\noindent
      \txc{red}{\Large  Recent comprehensive 3D numerical models of the ISM}
    }  
    \hfill
    \rotatebox{90}{
       \begin{tabular}{lcccccccc}
          \hline
                           &Korpi$^{\rm\bf(a)}$   &Avillez$^{\rm\bf(b)}$    &Hanasz$^{\rm\bf(f)}$  &Balsara$^{\rm\bf(c)}$   &Slyz$^{\rm\bf(d)}$  &Piontek$^{\rm\bf(d)}$  &Joung$^{\rm\bf(b)}$   &Gressel$^{\rm\bf(e)}$ \\[5pt] 
          \hline                                                                                                                                                                                  
          Domain, kpc      &$0.5\times0.5\times2$ &$1\times1\times5$--$ 20$ &$0.5\times1\times1.2$ &$0.2^{3}$               &$1.28^{3}$          &$0.2^{3}$              &$1\times1\times10$    &$0.5\times0.5\times4$ \\[5pt]
          Res, pc          &$8$                   &$1.25$--$10$             &$10$--100             &$0.78$--$1.56$          &$10$--$20$          &$0.78$--$1.56$         &$1.95$                &$2$--$8$              \\[5pt]
          Heating          &SN~II+SN~I            &SN~II+SN~I               &CR                    &SN                      &SN                  &--                     &SN~II+SN~I            &SN~II+SN~I            \\[5pt]
          Temp, K          &$10^2$--$10^8$        &$2\times10^2$--$10^6$    &$2\times10^2$--$10^6$ &$10^2$--   $10^8$       &$10^2$--$10^8$      &$10^{1.5}$--$m10^4$     &$2\times10^2$--$10^7$ &$1$--$10^7$           \\[5pt]
          Phases           &warm(W)+hot(H)        &cold(C)+W+H              &C+W+H                 &W+H                     &C+W+H               &C+W                    &C+W+H                 &C+W+H                 \\[5pt]
          MHD              &non-ideal             &hydro\& ideal            &non-ideal             &ideal                   &hydro               &ideal                  &hydro                 &non-ideal             \\[5pt]
          Diffusion        &hyper $\eta~\nu~\chi$ &--                       &$\eta$                &--                      &--                  &$\chi$                 &--                    &$\eta~\nu~\chi$       \\[5pt]
          Gravity          &stellar+halo          &stellar+halo             &stellar+halo          &--                      &self gravity        &linear z               &stellar+halo          &yes                   \\[5pt]
          Rotation         &differential          &--                       &rigid \& differential &--                      &--                  &differential           &--                    &differential          \\[5pt]
          Cosmic rays      &--                    &--                       &yes                   &--                      &--                  &--                     &--                    &--                    \\[5pt]
          Star forming     &--                    &--                       &--                    &--                      &yes                 &--                     &--                    &--                    \\[5pt]
          Duration         &100 Myr              &400 Myr                  &4.8 Gyr               &--                      &350 Myr             &2 Gyr                  &80 Myr                &1 Gyr                 \\[5pt]
          \hline
        \end{tabular}
    }
    \hfill
    \rotatebox{90}{
A summary of three dimensional simulated models
of galaxy features by: \cite{Korpi99}(a), \cite{Avillez07}(b) from 2000,}
    \rotatebox{90}{
\cite{Hanasz09a}(c) from 2004, \cite{Balsara04}(d), \cite{Slyz05}(e),
\cite{Piontek07a}(f) from 2005, \cite{Gressel08}(f).  }
\caption{}    
\label{table:sim}
\end{table}  

    \subsection{Korpi}
      \label{ssection:korpi}

        \citet*{Korpi99} constructed a comprehensive 3D model of a section of
the galactic disc using parameters typical of the solar neighbourhood. The
simulation included non-ideal MHD applied to a Cartesian grid of
$0.5\times0.5\times2$ kpc$^{3}$. The resolution was about 8$^3$ parsecs$^{3}$
(pc).

        A vertical gravitational field (\citet*{Kuijken89}) and a shearing box 
to replicate the differential rotation were added. Thus a vertically
stratified density profile for the interstellar medium (ISM) was applied.
Heating was applied via Supernovae (SNe) Type I and II distributed randomly
in time and space. Thermal conductivity, kinetic and magnetic viscosities 
were modelled and hyper viscosity applied to handle shocks.

        The model replicated a multi-phase medium with the warm and hot
phases in dynamic pressure equilibrium. The cold phase was not well
reproduced through lack of a thermally unstable cooling function and low
resolution. 

        The resultant ISM structure was highly differentiated in $Z$. Volume
filling factors for temperature varied with height. Correlations for mass and
velocity varied with temperature. The outward flow of the galactic fountain
appeared to be  described within the model, but the size of the model was
insufficient to recover the global fountain behaviour.
 
       In this study the warm phase was characterized by strong turbulent flow,
 but in the hot phase flows were on the scale of the remnant size enclosing the
 hot regions. This could be a feature of the numerical limits of the model.

    With this model the group also studied the evolution of super-bubbles
 (SBs) of SNe clusters and how this was affected by the strength of the
 magnetic field and turbulence. They found that the frequency of SB break outs
 from the disc to the halo was significantly increased in a turbulent regime
 compared to previous experiments in a homogeneous medium.
 
       The role of magnetic field strength in confining the bubbles was
reduced. This could be arise from the increased random element to the field
due to turbulence. Thus our expectation of the mixing between the disc and
the halo would be higher.
        
      \subsection{de Avillez}
        \label{ssection:avillez}
        
     \begin{figure}[hb]
      \label{av.fig}
      \begin{center}
       \includegraphics[width=\linewidth]{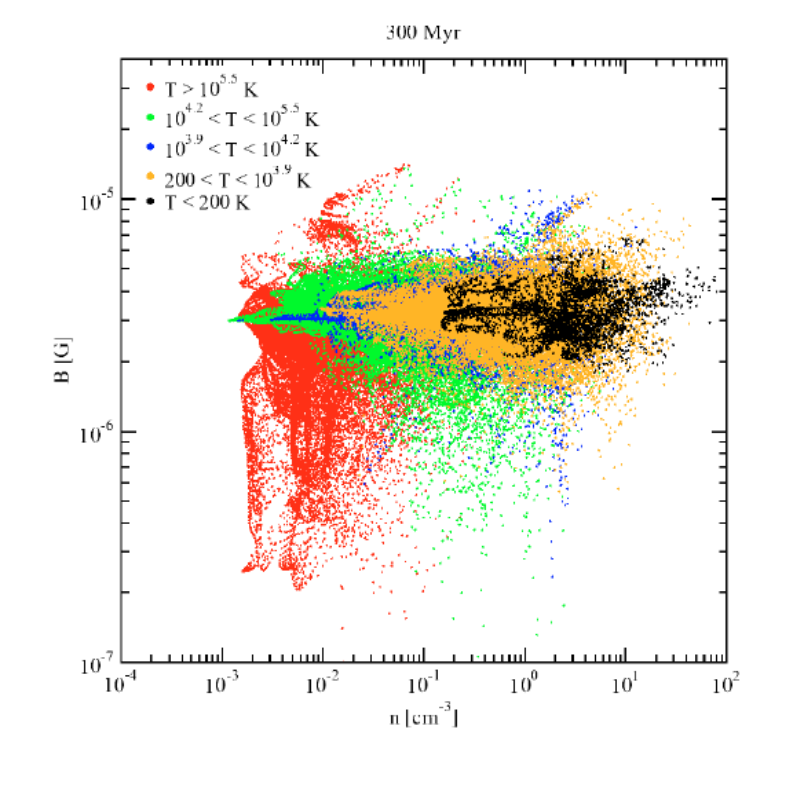}
       \caption{{\footnotesize{\cite{Avillez07} Scatter plot from study (2005)
of $B$ versus $\rho$ for $T ≤ 200$ (black), $200 < T ≤ 10^{3.9}$ (orange),
$10^{3.9} < T ≤ 10^{4.2}$ (blue), $10^{4.2} < T ≤ 10^{5.5}$ K (green), and
$T > 10^{5.5}$ K (red) regimes at 300 Myr of disc evolution. The points in the
plot are sampled at intervals of four points in each direction. Note that
during the evolution of the system the field strength broadened its
distribution spanning two orders of magnitude after 300 Myr.}}}
      \end{center}
     \end{figure}

        Using a model since 2000 \citet*{Avillez07} applied 3-D HD equations to
a Cartesian grid 1 kpc$^{2}$ by 8 kpc. They used adaptive mesh refinement
(AMR) with a resolution ranging between 1.25$^3$ and 10$^3$ pc$^3$. 
Also included was a vertical gravitational field and Type Ib$ + $Ic and
Type II SNe. 60\% were clustered in OB associations. Thermal conductivity
and kinetic viscosity were neglected, but a thermally stable radiative
cooling function was included. 

       The model evolved from hydrostatic equilibrium over a period of about
 65 Myr to a dynamically steady state. Distinct cold ($T\le 10^{3}$K),
warm ($10^{3} < T \le 10^{4}$K), warm ionized ($10^{4}<T\le 10^{5}$K) and
hot ($T > 10^{5}$K) phases were co-existing with varying scale heights.
The cold phase was restricted to thin irregular strips around the mid-plane.
Hot gas dominated above 1.5 kpc and warm gas in the disc below 500 pc. 

       The model was extended vertically to $\pm 10$ kpc between 2001 and 
2007. They identified bubbles and super-bubbles typically up to 120 pc across
and 200 pc high breaking out of the disc periodically. The cold sheet like
structures that evolved, and which resembled the fragments of SN remnants,
were in fact found not necessarily to be correlated with the bubble remnants,
but in fact appeared to be a more general effect of the turbulent
compressions and convection currents. 

       For a further adaption of the model to MHD an initial uniform 5.8$\mu$G
magnetic field in the azimuthal direction was included.  The horizontal
magnetized disc only briefly delayed the disc-halo cycle observed in the
HD simulations and a hot phase volume filling factor of between 17 to 21\%
resulted both with HD and MHD.

        Supersonic and super-Alfv{\'e}nic flows led to strong MHD shocks. 
A highly turbulent field evolved with a field strength ranging between
0.1 and 10$\mu$G, which was not strongly correlated to the density
fluctuations. This is illustrated in figure \ref{av.fig}. Magnetic
pressure dominated the coldest gas, ram pressure dominant above 200 K and
thermal pressure above $10^{5.5}$ K. In cold regions the relation 
$B \propto \rho^{\alpha}$ did not hold, with $\alpha$ in the range -0.006
to 0.085. 

     \subsection{Balsara}
       \label{ssection}

       \citet*{Balsara04} applied 3D ideal MHD to a
 (200 pc)$^{3}$ region with SN randomly distributed. They included cooling
\citep*{Raymond76}  and a constant diffuse heating term. Initial equilibrium
was achieved by balancing heating and cooling for some given isothermal 
uniform density.        The cooling considered could reproduce hot and warm phases, but not the  cold phase.

      They explored the results when varying the initial temperature and 
density and applying SN rates 1 and 100$\times$ the rate in the solar 
neighbourhood. They investigated resolution at about 1.5$^3$ and 0.75$^3$
pc$^3$. Stratification and differential rotation were not considered nor 
thermal conductivity and kinetic viscosity. 

      They found evidence of a fast dynamo sensitive to the SN rate. Over a
40 Myr simulation the field was amplified by two orders of magnitude. The 
growth rate was stronger in the high resolution, due to additional 
amplification available at smaller scales.

      Taking a turbulent scale from an SN remnant to be 30pc they compared energy growth at scales above and below this. They found that eventually 
the growth rates were independent of scale. 

      Growth rates however were sensitive to SN rates, increasing up to a 
threshold where the magnetic energy was quenched. The growth of the field
depended on the generation of helicity from the SN shocks interacting with
the warm ISM. This was suppressed as the ISM became saturated by hot
remnants. This effect had a critical rate somewhere between 12 and 40 times
the galactic rate. There was also some increased amplification if the SN 
rate was more intermittent and field growth was most vigorous where the ISM
had lower temperature and higher density.

     \subsection{Hanasz}
       \label{ssection:hanasz}

     \begin{figure}[htp]
      \label{fig:hanasz}
      \begin{center}
       \includegraphics[width=0.6\paperwidth]{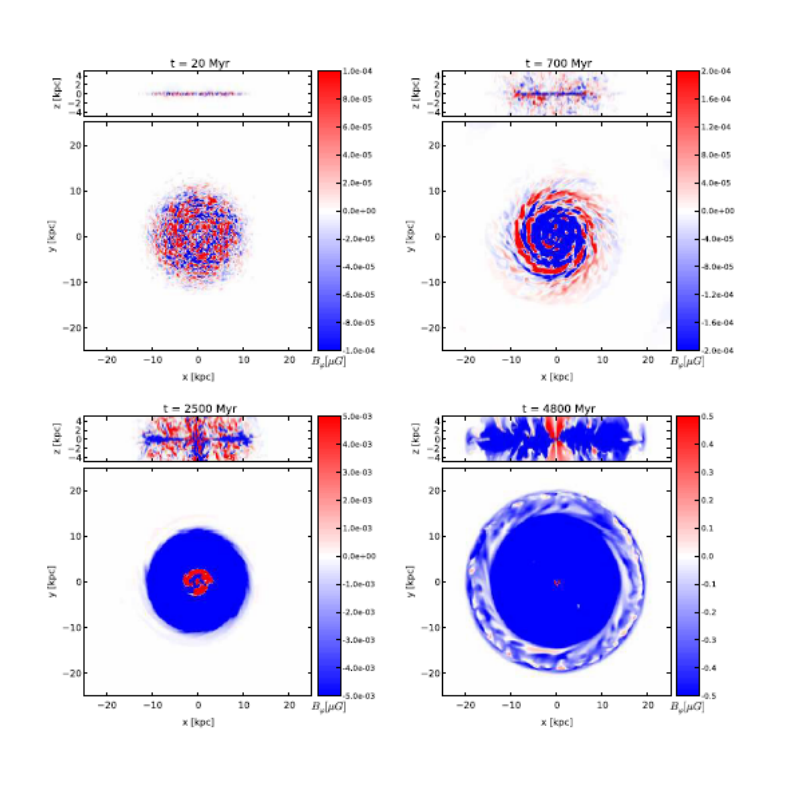}
       \caption{{\footnotesize{Distribution of toroidal magnetic field from
\cite{Hanasz09a}. Unmagnetized regions are white, while positive and
negative toroidal fields are shaded red and blue, respectively. Note that
the color scale in magnetic field maps is saturated to enhance weaker
magnetic field structures in the disc peripherals. The maximum magnetic
field strengths are $5.9\times10^{-4},4.4\times10^{-3},1.5$ and 29$\mu$G at
$t=0.02,0.7,2.5~\rm{and}~ 4.8$ Gyr respectively. }}}
      \end{center}
     \end{figure}

       Form 2002 \citet*{Hanasz09a} investigated dynamo action with rigid
rotation on a magnetized galactic disc. A uniform azimuthal field
was applied to a rotating Cartesian domain $600 \times 1800 \times 600$ pc 
with resolution $5 \times 20 \times 2.5$ pc. They solved isothermal resistive
MHD equations (neglecting kinetic viscosity).
       
       In 2004 they included cosmic rays (CR) randomly injected at spherical
locations representing SN remnants. Differential rotation was now included in
a $0.5\times 1 \times 1.2$ kpc$^{3}$ Cartesian grid, with a resolution of 
10$^3$ pc$^3$. The gravitational potential was as described by \citet*{ferriere01} and the cosmic ray diffusion tensor that of \citet*{ryu03}

       They observed an exponential growth in the magnetic field over a
period of about 1000 Myr. Growth in the vertical component dominated, 
indicating that the CR buoyancy was important to the dynamics. Growth in
the azimuthal direction reached 1400$\times$ the seed field amplitude and 400
 $\times$ in the radial direction. 

      Growth in the ideal regime onset rapidly after 250 Myr, due to
magnetorotational instability (MRI). They argued that MRI amplification is 
suppressed when comparing the resistive MHD to ideal MHD due to magnetic 
reconnection in the thin current sheets generated by compressions in the
field.

      They observed that the fastest growing dynamo model resulted from
periodic switching on and off the SN activity. Increased SN activity enhanced
vertical field lines, increasing the loss of CR to the halo. Reduced activity
supported their horizontal realignment thus enabling the amplitude of the
field to grow. So the irregularity of the SN distribution can add to the
dynamo.

       \cite{Hanasz09a} investigated the source of the galactic magnetic seed
field using their CR-driven dynamo model. Their domain was $50^2\times10$
kpc$^3$ with resolution 100$^3$ pc$^3$. They started with an isothermal,
vertically and radially stratified density, with a gravity potential and
rotational shear, neglecting the central bulge.
 
       Into an unmagnetized ISM they randomly exploded SN remnants. Some
10\% of these included a randomly oriented weak dipolar magnetic field and
CR equivalent to 10\% of the typical kinetic energy of an SN remnant.
Over a 4.8 Gyr simulation they produced a magnetic field, which became 
efficiently ordered within 2 Gyr, aligned along the spirals and including 
field reversals. The vertical field evolves an x-shaped configuration. In
their movies, vertical outflows could be seen, which transported the counter
polarity or helicity out of the system (Figure \ref{fig:hanasz}, small 
inserts). This seemed to be crucial for dynamo action.

       The number and orientation of the spiral configurations is sensitive
to the CR diffusion coefficients. They could not be derived from first
principles without smaller scales, so the assumed value for diffusivity
was large and unrealistic by necessity.  Snap shots of the 
configuration are shown in Figure \ref{fig:hanasz}

     \subsection{Slyz}
       \label{ssection:slyz}

       In their 2002 2D model \citet*{Slyz05} investigated the characteristics
of  the spiral arms through isothermal HD. They used a 23$^2$ kpc$^{2}$ with a resolution of 115$^2$ pc$^2$, including kinematic viscosity and applying a centrifugal and gravitational potential. 

       They varied the temperature, via the sound speed, the density profiles 
and rates of galactic rotation. Given the typical sound speed throughout the
galaxy is far exceeded by the angular velocity of the gas relative to the 
spiral arms, it might not have been expected to have much effect the 
structure. They found that the non-axisymmetric pattern of the spiral galaxies
diminished with increasing sound speed. At levels above 25 to 30 km s$^{-1}$
they faded away. As sound speeds reduced the mass in the spiral arms 
increased. 

      They also found that the inner galaxy was more sensitive to changes in
the fraction of galactic mass in the disc. As the disc fraction increased,
velocities at the centre increased  massively, whilst the structure in the
outer regions was fairly insensitive to the disc fraction. 

     Subsequently \cite{Slyz05} conducted a more localized analysis of star
formation and SN feedback with a 3D hydrodynamical model. They used a 
1.28$^{3}$ kpc$^{3}$ Cartesian grid together with a particle mesh to track 
the stellar mass. The model included a thermally unstable cooling function 
and self-gravity. Stratification and any external gravity were not included.

     A random Gaussian velocity perturbation was applied at the start onto
an ISM of uniform density and temperature. They found that the regime 
saturated to a multi-phase medium, with velocity dispersion generally 
settling to levels consistent with the temperature of the medium: 15 km
s$^{-1}$, 30 and 75 for cold ($<2000$ K), warm ($<10^{5}$ K) and hot
($<4\times 10^{6}$ K) respectively. Velocities in the very hot ($>4\times
10^{6}$ K) are far more dispersed, with up to 500 km s$^{-1}$ observed. 
These violent flows strip away elements of the cold dense star forming 
regions as they interact.

     The distribution of mass was also a feature of the ISM temperature. 
The bulk of the volume was filled with hot diffuse medium of below $10^{-3}$ 
atoms cm$^{3}$. The bulk of the mass was restricted to dense cold 
filamentary or sheet-like or worm-like structures accounting for a very small
portion of the volume. In the absence of SN shocks the probability density functions (PDF) for the ISM were well described by the log normal. However
with SN feedback a much higher incidence of increased density regions were
observed, suggesting that star formation rates might be enhanced in a 
turbulent ISM.

    \subsection{Mac Low}
      \label{ssection:maclow}

     Continuing the work from \cite{Balsara04}, \citet*{MacLow05} studied
the pressure distributions in the turbulent ISM with slightly differing 
results to those of \cite{Slyz05}. They applied ideal MHD equations in a
(200 pc)$^{3}$ domain, with 3.12, 1.56 and 0.83 parsec resolution, including
a radiative cooling function and uniform uv-heating, but excluding gravity. 

      They randomly injected SN remnants, at 1 and 4 times the solar
neighbourhood rate and analysed the evolution of the pressure distribution.
They found the distribution of the pressure anomalies was independent of
the orientation of the magnetic field due to the super-Alfv{\'e}nic flows. 
Typically the hot gas formed discrete clumps, which were enclosed  by warm
or cold gas. The distribution of the colder dense clusters, tended not to be
closely aligned directly to remnant structures. These formed on scales of
dozens of parsecs and appeared to result from the general effects of large
scale turbulence.

     The presence of dense isolated filaments was more effective at breaking
the momentum of remnants than equivalent mass more smoothly distributed
and even with 4 times the SN rate hot zones remained unconnected and warm
gas remained the largest volume filling factor. Even with SN driving they
found the PDFs of the pressure were well described by the log normal and not
a power law. The dispersion range was an order of magnitude away from
the mean, far from equilibrium.

     \subsection{Piontek}
       \label{ssection}

       The turbulence in the galaxy is not adequately accounted for by
SN activity alone so in 2005 \citet*{Piontek07a} modelled MRI-driven
turbulence in the warm/cold phases of the ISM. They apply ideal MHD to a
$(200 \rm{pc})^{3}$ domain with resolution of 0.78 and 1.56 pc and realistic
heating and cooling. They model differential rotation for the solar 
neighbourhood.
       The growth of turbulence was studied in a medium in uniform density 
and pressure and also one in which randomly distributed cold spherical
clouds were included amongst the warm ISM. The MRI instability rapidly
dominated and the saturated state was independent of the initial state.
       They studied the relationship between the MRI driven turbulence
and the two-phase thermal structure derived from the cooling function
by comparison to MRI is a single phase environment.
     
       They found that the gas separated into two phases with peaks at
100 K and 8000 K, with an increasing spread of gas lying outside the
equilibrium as turbulence saturated. The relative proportions of gas
in the two phases was dependent on mean density, with the proportion of
mass in the cold phase increasing substantially as the mean density 
increased.
 
        Pressures were in reasonable equilibrium even in two phases
without turbulence, but in the saturated state the pressure dispersion
increased five fold.
        
        Magnetic field growth 0.26$\mu$G to about 2.5$\mu$G was not 
sensitive to density nor phase.  
        Away from SN activity MRI could certainly be a factor in the
turbulent structure. Extrapolating their results for mean densities
 typical in the inner galaxy, $R<10$, kpc they would predict turbulent 
velocity scales of about 4$\kms$, which could be a significant 
component.     

        \cite{Piontek07a} included a vertical gravity profile, which 
now induced a vertical stratification in both density and temperature.
The turbulent mixing of the warm ISM had a significantly higher scale
height than the cold, which now was more clustered than in the earlier
model. 
        This combined with MRI to increase the magnetic pressure and 
they argued that the MRI could be responsible for suppressing star
formation in the less dense outer region of the mid plane, even given
the presence of cold clouds.

      \subsection{Joung}
        \label{ssection:joung}

        From 2006 \citet*{Joung09} conducted a HD study in a vertically stratified 
ISM including stellar and halo gravity and realistic diffuse heating
and cooling plus SN turbulence. The vertical profile of the heating was
realistically motivated. They neglected thermal conduction and
kinetic viscosity, self-gravity and differential rotation. 

        They found the scale height of the galactic fountain to be several
kpc. The volume filling factor near the mid-plane of the hot gas was 
inversely related to the magnitude of the diffuse heating, which they tested
over a range up to 4 times that of \citet*{Wolfire95}. This was due to
a shift in the thermal equilibrium position to exclude the cold phase and 
increase the volume component of the warm gas. 

        The vertical density profile was out of agreement with observations
which they attributed to lack of pressure, in the absence of magnetic and 
cosmic ray pressure. 
        
        In this study and also from 2009 they find that neither 
the density nor the kinetic energy wavelengths follow the Kolmogorov power 
law, but instead are far more widely dispersed.

      \subsection{Gressel}
       \label{ssection:gressel}

     \begin{figure}[htp]
      \label{fig:gressel}
      \begin{center}
       \includegraphics[width=0.6\paperwidth]{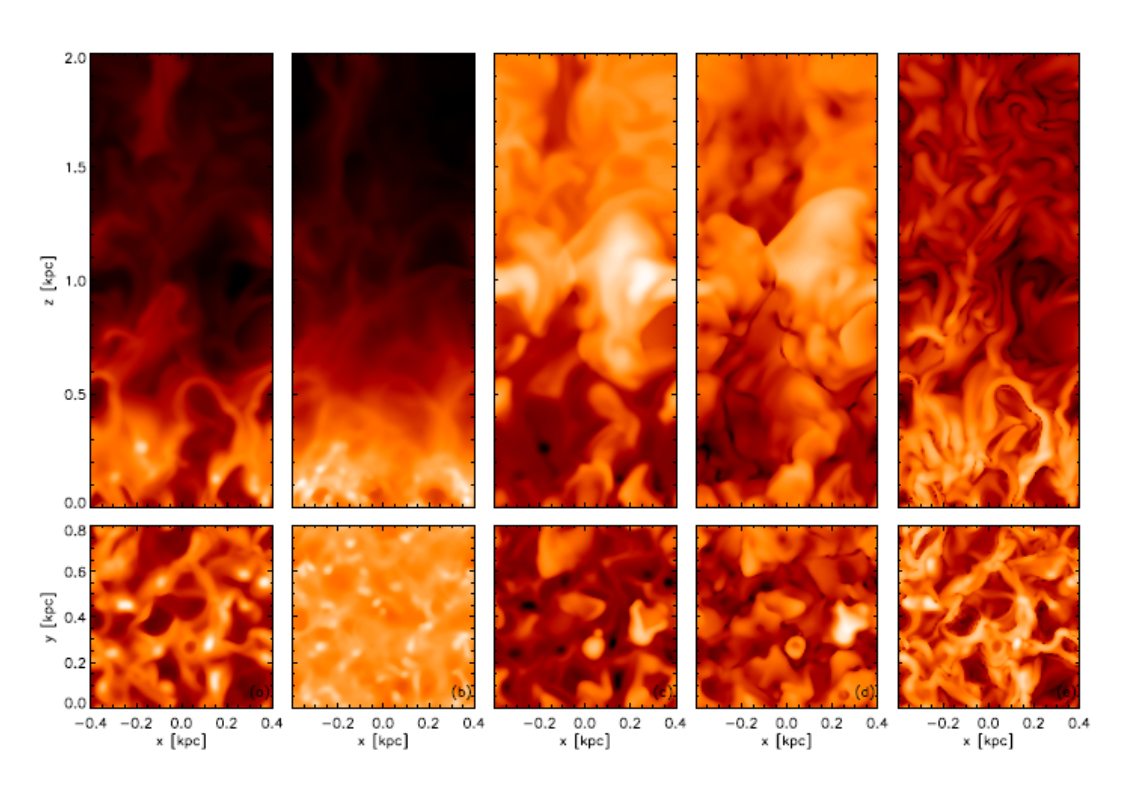}
       \caption{{\footnotesize{Vertical slices from above the mid-plane
(\it{upper panels}) and horizontal mid-plane slices (\it{lower panels})
after $t=112$ Myr from
\cite{Gressel08}. From left to right [in logarithmic scales]: number
density[-4.82,1.10]cm$^{-3}$,column density [17.34,21.55]cm$^{-2}$,
temperature[2.06,7.20]K, velocity dispersion[-0.15,2.77]km s$^{-1}$ \&
magnetic field strength[-4.16,0.34]$\mu$G. }}}
      \end{center}
     \end{figure}

        \citet*{Gressel08} conducted a non-ideal 3D
MHD simulation using the NIRVANA code. Their domain was $0.8^2\times 4$
kpc$^{3}$ and included realistic cooling, thermal conductivity and 
kinematic viscosity and rotational shear. A snapshot from their simulation
is shown if figure \ref{fig:gressel} to illustrate the resolution and 
structure obtained through such simulations.

        They tested a variety of SN rates ($1/4$ to $1\times\sigma_{0}$) 
and rotational frequency (1 to $8\times\Omega_{0}$), where $\sigma_{0}$ 
and $\Omega_{0}$ denote the observed rates in the solar neighbourhood, and
measured the resulting dynamo parameters. 

        A result common to each run, was that the velocity dispersion 
increased with height up to about 1 kpc, even though the peak of SN 
activity was much lower. They found that turbulent pumping was directed
inward, but balanced by a outward wind of comparable magnitude.

        They found the dynamo not to be sensitive to the SN rates considered,
1, 2 \& $4\times$ the observed rates. However the dynamo increased only with
galaxy rotation rates 2 \& $4\times$ the observed rate. They did not rule out 
however that these conclusions could also be sensitive to the assumed gravity
and density profiles. 
        They found no dynamo in the models without rotational shear.

   \section{Summary}
    \label{section:summary}
 
     At the highest level of resolution we have had the study of \cite{Balsara04},
\cite{Piontek07a} and \cite{MacLow05} in $200^3$ pc$^{3}$ domains. Too small to 
include large scale structure, such as density stratification or spiral arms, 
these studies were able to relate the relationship of turbulence from SN or MRI,
with dynamo amplification.

     They were able conclude that MRI could be a significant source of turbulence
in the galaxy and in particular in the outer galaxy away from SN activity. The
presence of cold dense filaments, they demonstrated, were not closely correlated
to the dynamics of the SN remnants. Instead they appeared to be a product of
generalised turbulence and gravitational instability.

     Finally they found dynamo amplification rate was sensitive to the level SN
rate up to an upper bound of between 12 \& $40\times$ the solar neighbourhood rate.
Amplification of the magnetic field was also increased by greater intermittency
in the SN rate. Independent of SN rate the dynamo is quenched once the ISM is saturated by hot
   
     Losing resolution, but gaining structure \cite{Korpi99}, \cite{Avillez07},
 \cite{Slyz05}, \cite{Joung09} and \cite{Gressel08} constructed models 
incorporating the density stratification of the ISM near the plane of the 
galactic disc. 

     Unlike the smaller models, the ISM does not become saturated by SN activity.
The stratified structure permits the disc to expand and relax as the SN rate 
fluctuates in time and spatial distribution. This relieves the quenching effect
of turbulent saturation on the magnetic dynamo.

     However results for the dynamo have been mixed. \cite{Korpi99} and
\cite{Gressel08} did not find a dynamo with solar neighbourhood model 
parameters, and \cite{Gressel08} only found the dynamo with differential
rotation 2 and $4\times$ the galactic rate. Both these models used non-ideal
MHD, where magnetic reconnection may dampen the dynamo in the critical cold
dynamic filaments. The numerical values for resistivity are of course much
higher than the realistic values observed in the ISM. \cite{Avillez07} did
not investigate the dynamo but included a 5$/mu$G magnetic field using ideal
MHD within their later work. 

     With a vertical range of 20 kpc \cite{Avillez07} were able to include
elements of the galactic fountain and without boundary mass losses, could
sustain simulations over 400 Myr. They found that an excess of 200 Myr was 
required to completely saturate and establish an equilibrium state for the
disc-halo cycle. They identified the scale height of the disc-halo interface
at about $\pm$1.5 kpc. The additional height was required to allow the hot
gas in the halo to cool and rain back to the disc.
     
     Despite outflow boundary conditions, and hence mass losses over the 
duration of the simulation, \cite{Gressel08} were able to sustain their
simulations over 1 Gyr. Once the turbulent state had evolved they found the
boundary flows were minimal, so much of the dynamical cycle could be 
contained within a domain height of $\pm2$ kpc.
 
     On the grand scale \cite{Hanasz09a} were able to produce a familiar
model of the spiral structure of the galaxy. Restricted to isothermal models,
they found the structure of the spirals sensitive to the CR diffusion. In 
addition the 2D study of \cite{Slyz05} found the structure was also dependent
on the temperature of the ISM, as parameterized by sound speed. 

     These models have informed our understanding of how the ISM behaves on different scales across a range of parameters. It is also clear that the 
interaction of temperature, density, magnetic field and cosmic rays combined
are significantly altered in the absence of any one of these with unexpected
consequences. 

     The challenge is to successfully incorporate all these elements
over a number of physically meaningful scales. Global models of galaxies are
indispensable, with sufficiently high resolution to include the small-scale 
physics. To resolve SNe remnants requires a resolution under 5$^3$ pc$^3$. 
For the random field dynamo we may require scales $<2^3$ pc$^3$. The gravitational field (and self-gravitation of the gas) needs to be included, 
the latter requiring scales below 1pc. Density waves are expected to interact 
with the magnetic fields and need to be considered. Arm and inter-arm modelling
would require domains extending a few kpc in the galactic plane. 
To include even the lowest reach of the halo, would require a vertical range
of at least 4 kpc. 

\bibliographystyle{apsfred-nameyear}      
\bibliography{fred}   

\end{document}